\documentclass[11pt]{article}
\usepackage{amsmath}
\usepackage{amsfonts}
\usepackage{amssymb}
\usepackage{theorem}
\newtheorem{theorem}{Proposition}
\newtheorem{lemma}{Lemma}
\newcommand{\dbar}{\bar{\partial}}

\newcommand{\be}{\begin{equation}}
\newcommand{\ee}{\end{equation}}
\newcommand{\bea}{\begin{eqnarray}}
\newcommand{\eea}{\end{eqnarray}}
\newcommand{\beaa}{\begin{eqnarray*}}
\newcommand{\eeaa}{\end{eqnarray*}}

\newcommand{\nn}{\nonumber}
\begin{document}
\title
{On the heavenly equation hierarchy and its reductions}
\author{
L.V. Bogdanov\thanks
{L.D. Landau ITP, Kosygin str. 2,
Moscow 119334, Russia}~~and
B.G. Konopelchenko\thanks
{Dipartimento di Fisica dell' Universit\`a di Lecce
and Sezione INFN, 73100 Lecce, Italy}}
\maketitle
\begin{abstract}
Second heavenly equation hierarchy is considered using the framework
of hyper-K\"ahler hierarchy developed by
Takasaki \cite{Takasaki,Takasaki1}. Generating equations for the hierarchy are introduced,
they are used to construct generating equations for reduced hierarchies. General
$N$-reductions, logarithmic reduction and rational reduction for one of
the Lax-Sato functions are discussed. It is demonstrated that rational reduction
is equivalent to the symmetry constraint.
\end{abstract}
\section{Introduction}
Plebansky second heavenly equation \cite{Plebanski},
having its origin in general relativity,
has attracted a lot of interest both from the viewpoint of integrability and relativity.
It has been intensively studied using different techniques
(see e.g. \cite{Penrose,dun1,dun2,FP,FerK,dun3,MS}).

In the work \cite{heav} we have developed a $\dbar$-dressing scheme applicable to second
heavenly equation. A very important role was played by a kind of Hirota bilinear
identity, which leads to the introduction of the function $\Theta$ (analogue
of the $\tau$-function for heavenly equation hierarchy) and produces the hierarchy
in the form of addition formulae (generating equations) for $\Theta$. This identity
also has its natural place in the framework of hyper-K\"ahler hierarchy developed
by Takasaki \cite{Takasaki,Takasaki1},
who demonstrated that it is equivalent to the Lax-Sato equations of the
hierarchy. Here we will use this framework to study the
reductions of the heavenly equation hierarchy and its symmetry constraints. The ideas and logic
of this work are very close to the works \cite{BKA,dtau,dconstr}, where dispersionless
hierarchies were considered.
\section{Heavenly equation hierarchy}
First we intoduce the principal objects and  notations.
We start from two formal Laurent series in $z$,
\begin{gather}
S^1=\sum_{n=0}^\infty t^1_n z^{n}+\sum_{n=1}^\infty S^1_n(\mathbf{t}^1,\mathbf{t}^2)z^{-n},
\label{form1}
\\
S^2=\sum_{n=0}^\infty t^2_n z^{n}+\sum_{n=1}^\infty S^2_n(\mathbf{t}^1,\mathbf{t}^2)z^{-n},
\label{form2}
\end{gather}
where the variables $t$ are considered independent and $S^1_n$, $S^2_n$ are dependent variables.
We denote $x=t^1_0$, $y=t^2_0$,
$
\mathbf{S}=
\left(
\begin{array}{c}
S^1\\
S^2
\end{array}
\right),
$
introduce the Poisson bracket
$\{f,g\}:=f_x g_y-f_y g_x$ and the projectors $(\sum_{-\infty}^{\infty}u_n z^n)_+
=\sum_{n=0}^{\infty}u_n z^n$,
$(\sum_{-\infty}^{\infty}u_n z^n)_-=\sum_{-\infty}^{n=-1}u_n z^n$.

Heavenly equation hierarchy is defined by the relation (see \cite{Takasaki}, \cite{heav})
\be
(d S^1\wedge d S^2)_-=0,
\label{analyticity0}
\ee
playing a role similar to the role of the famous Hirota bilinear identity for KP hierarchy.
This relation is equivalent to the Lax-Sato form of the hierarchy.
\begin{theorem}
The identity
(\ref{analyticity0})
is equivalent to the set of equations
\bea
&&
\partial^1_n\mathbf{S}=-\{(z^n S^2)_+,\mathbf{S}\},
\label{Hi1}
\\
&&
\partial^2_n\mathbf{S}=\{(z^n S^1)_+,\mathbf{S}\},
\label{Hi2}
\\
&&
\{S^1,S^2\}=1.
\label{bracket}
\eea
\label{formHEH}
\end{theorem}
The proof of this statement is given in \cite{Takasaki}
for the general hyper-K\"ahler hierarchy
(second heavenly equation hierarchy is its two-component special case). 
We will not reproduce
the complete proof, but will just illustrate some ideas, which will be useful later.
It is possible
to prove that (\ref{analyticity0}) implies Lax-Sato equations using the following
statement.
\begin{lemma}
Given identity (\ref{analyticity0}),
for arbitrary first order operator $\hat U$,
$$
\hat U\mathbf{S}
=\sum_i (u^1_i(z,\mathbf{t}^1,\mathbf{t}^2)\partial^1_i
+u^2_i(z,\mathbf{t}^1,\mathbf{t}^2)\partial^2_i)
\mathbf{S}
$$
with `plus' coefficients ($(u^1_i)_-=(u^2_i)_-=0$),
the equality
$(\hat U\mathbf{S})_+=0$ implies that $\hat U\mathbf{S}=0$.
\label{operator}
\end{lemma}
\textbf{Proof}
First, (\ref{analyticity0}) implies that
$$
\{S^1,S^2\}_-=0,
$$
and, using (\ref{form1}), (\ref{form2}), we get
$$
\{S^1,S^2\}=\{S^1,S^2\}_+=1.
$$
Then, using (\ref{analyticity0}) we obtain that
\beaa
\hat U S^1\cdot S^2_x - \hat U S^2\cdot S^1_x=0,\\
\hat U S^1\cdot S^2_y - \hat U S^2\cdot S^1_y=0.
\eeaa
If $\hat U\mathbf{S}\neq0$, then we come to the conclusion that vectors
$\mathbf{S}_x, \mathbf{S}_y$ are linearly dependent, then
$\{S^1,S^2\}=0$, and we come to a contradiction.
\hfill$\square$\\

The proof of  Proposition \ref{formHEH} (sufficient condition) is then straightforward,
one should just check directly that
\beaa
&&
(\partial^1_n\mathbf{S}+\{(z^n S^2)_+,\mathbf{S}\})_+=0,
\\
&&
(\partial^2_n\mathbf{S}-\{(z^n S^1)_+,\mathbf{S}\})_+=0.
\eeaa

\subsection*{Function $\Theta$}
Identity (\ref{analyticity0}) leads also to the introduction of an analogue
of the $\tau$-function in terms of closed one-form.
\begin{theorem}
The one-form
\be
\theta=\frac{1}{2\pi\mathrm{i}}\mathrm{Res}_\infty(S^2_- d S^1_+ - S^1_- d S^2_+)
\label{1-form}
\ee
is closed.
\end{theorem}
\textbf{Proof}
Identity (\ref{analyticity0}) implies that
$$
d\theta=\frac{1}{2\pi\mathrm{i}}\mathrm{Res}_\infty(dS^2_-\wedge d S^1_+ - dS^1_- \wedge d S^2_+)=0.
$$
\hfill$\square$\\

Similar to the work \cite{Takasaki},
we define a $\tau$-function $\Theta(\mathbf{t}^1,\mathbf{t}^2)$
for heavenly equation hierarchy
through closed one-form (\ref{1-form}) by the relation $d\Theta=\theta$. Introducing
vertex operators $D^1(z)=\sum_{n=0}^\infty z^{-n-1}\partial^1_n$,
$D^2(z)=\sum_{n=0}^\infty z^{-n-1}\partial^2_n$, it is easy to demonstrate that
\be
S^1_-(z)=-D^2(z)\Theta, \quad S^2_-(z)=D^1(z)\Theta.
\label{repr}
\ee
Substituting this representation into (\ref{bracket}), we get the equation
\be
D^2(z)\Theta_x-D^1(z)\Theta_y-\{D^1(z)\Theta,D^2(z)\Theta\}=0
\label{heav1}
\ee
The first nontrivial order of expansion of this equation at $z\rightarrow \infty$
gives exactly the heavenly equation
\be
\Theta_{ty}-\Theta_{\tilde t x}-\Theta_{xy}^2+\Theta_{xx}\Theta_{yy}=0,
\label{heavenly}
\ee
where $t=t^1_1$, $\tilde t=t^2_1$.

Identity (\ref{analyticity0}) gives also a general set of
addition formulae (generating equations in terms of vertex
operators) for $\Theta$ \cite{heav},
\begin{multline*}
\frac{1}{z'-z}D^1(z'')(D^1(z')-D^1(z))\Theta-
\frac{1}{z''-z}D^1(z')(D^1(z'')-D^1(z))\Theta
\\
=D^1(z'')D^2(z)\Theta\cdot D^1(z')D^1(z)\Theta
-D^1(z'')D^1(z)\Theta\cdot D^1(z')D^2(z)\Theta, \qquad
\end{multline*}
\begin{multline}
\frac{1}{z''-z}D^2(z')(D^2(z'')-D^2(z))\Theta-
\frac{1}{z'-z}D^2(z'')(D^2(z')-D^2(z))\Theta
\\
=D^2(z')D^2(z)\Theta\cdot D^2(z'')D^1(z)\Theta
-D^2(z')D^1(z)\Theta\cdot D^2(z'')D^2(z)\Theta, \label{add}
\end{multline}
\begin{multline*}
\frac{1}{z'-z}D^2(z'')(D^1(z')-D^1(z))\Theta-\frac{1}{z''-z}D^1(z')(D^2(z'')-D^2(z))\Theta\\
=D^1(z')D^1(z)\Theta\cdot
D^2(z'')D^2(z)\Theta-D^1(z')D^2(z)\Theta\cdot
D^2(z'')D^1(z)\Theta. \qquad
\end{multline*}
Expansion of these equations into powers of parameters $z$, $z''$,
$z''$ generates partial differential equations for $\Theta$
of the heavenly equation hierarchy.
\subsection*{Generating Lax-Sato equations}
We introduce also generating equations for the Lax-Sato form of the hierarchy,
\bea &&
(z'-z)D^1(z')\mathbf{S}(z)=-\{S^2(z'),\mathbf{S}(z)\} \label{Gen1},
\\
&&
(z'-z)D^2(z')\mathbf{S}(z)=\{S^1(z'),\mathbf{S}(z)\},
\label{Gen2}
\eea
which are equivalent to the set of equations (\ref{Hi1}), (\ref{Hi2}), (\ref{bracket})
(that can be checked directly or using Lemma \ref{operator}).

It is interesting to note that (\ref{Gen1}), (\ref{Gen2}) imply
the following symmetric expressions for Poisson brackets:
\beaa
&&
\{S^1(z'),S^2(z)\}=1+(z'-z)D^2(z')D^1(z)\Theta,
\\
&&
\{S^1(z'),S^1(z)\}=(z-z')D^2(z')D^2(z)\Theta,
\\
&&
\{S^2(z'),S^2(z)\}=(z-z')D^1(z')D^1(z)\Theta.
\eeaa
\section{Reductions}
\subsection*{General $N$-reductions}
We will discuss first the properties of general reduction, when
one of the functions $S^1_-$, $S^2_-$ depends on $N$ independent
functions of times (i.e., only $N$ coefficients of expansion in
$z^{-1}$ are independent). Reductions of this type were studied a lot
in dispersionless case (see e.g. \cite{YK}, \cite{GT}).
\begin{theorem}
Following three statements are equivalent:\\
1)
\be
S^1_-(z,\mathbf{t}^1,\mathbf{t}^2)
=S^1_-(z,f_1(\mathbf{t}^1,\mathbf{t}^2),\dots,f_N(\mathbf{t}^1,\mathbf{t}^2)),
\label{Nfunctions}
\ee
2)
\bea
\partial^2_N S^1(z,\mathbf{t}^1,\mathbf{t}^2)-
\sum_{i=0}^{N-1}\phi_i(\mathbf{t}^1,\mathbf{t}^2)
\partial^2_i S^1(z,\mathbf{t}^1,\mathbf{t}^2)=0,
\label{linear}
\eea
3) $\frac{S^1_y}{S^1_x}$ is a rational function with $N$ poles,
\be
\frac{S^1_y}{S^1_x}=\sum_{i=1}^N\frac{u_i}{z-v_i},
\label{ratio}
\ee
where $f_i$, $\phi_i$, $u_i$, $v_i$ are some functions of times.
\label{Nred}
\end{theorem}
\textbf{Proof}
$1\Leftrightarrow 2$  is evident and it is
not connected with equations of the hierarchy; $1\Rightarrow 2$ requires some linear
algebra, and $2\Rightarrow 1$ is proved by the method of characteristics.
The absence of minus projector in (\ref{linear}) (in contrast with (\ref{Nfunctions}))
is connected with the fact that $S^1$  is of the form (\ref{form1}) and
$\partial^2_i S^1_-=\partial^2_i S^1$.

$2\Rightarrow 3$ Using equations of the hierarchy (\ref{Hi2}),
one obtains
$$
\frac{\partial^2_n S^1}{S^1_x}={H^2_n}_x\frac{S^1_y}{S^1_x}-{H^2_n}_y,
$$
where $H^2_n=(z^n S^1)_+$. Substituting these expressions to relation (\ref{linear})
divided by $S^1_x$, one gets
\be
\frac{S^1_y}{S^1_x}=\frac{{H^2_N}_y-\sum_{i=0}^{N-1}\phi_i{H^2_i}_y}
{{H^2_N}_x-\sum_{i=0}^{N-1}\phi_i{H^2_i}_x},
\ee
that is evidently a rational function with $N$ poles.

$3\Rightarrow 2$ Using equations of the hierarchy and formula (\ref{ratio}),
we come to the conclusion that all ratios $\frac{\partial^2_n S^1}{S^1_x}$
are rational functions in $z$ with $N$ poles in the same points, 
that implies relation (\ref{linear}).
\hfill$\square$\\

A short comment on the Proposition \ref{Nred}.
Formula (\ref{Nfunctions}) gives a standard definition of
$N$-reduction similar to the dispersionless case (see e.g. \cite{YK}, \cite{GT}).
Equivalent formulation (\ref{linear}) suggests invariance of the hierarchy under the action
of some vector field and it is probably useful for geometric interpretation of
$N$-reduction. And finally, statement 3 gives analytic characterization of the reduction
in terms of Lax-Sato functions. This statement implies also that all ratios
$\frac{\partial^2_n S^1}{\partial ^1_m S^1}$ are rational functions of $z$.

Similar statements are also known in the dispersionless case \cite{fer}.

\subsection*{Generating equations for reduced hierarchy}
To obtain linear equations of the reduced hierarchy, we use (\ref{ratio})
to express $S^1_y$ through $S^1_x$
in generating Lax-Sato equations,
\bea
&&
(z-z')D^1(z')S^1(z)
\nn\\&&\qquad
=\left(S^2_x(z')\sum_i\frac{u_i}{z-v_i} - S^2_y(z')\right)\partial_x S^1(z)
=U^1\partial_x S^1(z),\qquad
\label{linred1}
\\
&&
(z'-z)D^2(z')S^1(z)
\nn\\&&\qquad
=\left(S^1_x(z')\sum_i\frac{u_i}{z-v_i} - S^1_y(z')\right)\partial_x S^1(z)
=U^2\partial_x S^1(z),
\label{linred2}
\\
&&
\partial_{y}S^1=\sum_{i=1}^N\frac{u_i}{z-v_i}\partial_x S^1=V\partial_x S^1.
\label{linred0}
\eea
Compatibility conditions for these equations are
\bea
&&
(z-z')D^1(z')V-\partial_y U^1 +VU^1_x-V_xU^1=0,
\label{compgen1}
\\
&&
(z'-z)D^2(z')V-\partial_y U^2 +VU^2_x-V_xU^2=0.
\label{compgen2}
\eea
First, both equations (zero order term at $z=\infty$) give an important relation
\be
\Theta_{yy}=-\sum_i u_i
\label{Thetared}
\ee
connecting $S^1,S^2$ with $u_i$,
\be
S^1(z')=S^1_+(z')+D^2(z')\Theta,\quad
S^2(z')=S^2_+(z')-D^1(z')\Theta
\label{uconnect}
\ee

Considering equation (\ref{compgen1}) at $z=v_j$, one obtains a system
\bea
&&
(z'-v_j)D^1(z')u_j-u_jD^1(z')v_j
\nn\\&&\qquad
=-(S^2_x u_j)_y+2S^2_{xx}(z')
\sum_{i(i\neq j)}\frac{u_i}{v_j-v_i} -(u_jS^2_{xy}-u_{jx}S^2_y)
\nn\\&&
(z'-v_j)D^1(z')v_j=S^2_{xx}u_j +(S^2_y v_{jx}-S^2_xv_{jy}).
\label{nred1}
\eea
Taking into account expressions (\ref{uconnect}), this is a closed (2+1)-dimensional
system of equations generating $t^1_n$ flows of reduced hierarchy.

Equation (\ref{compgen1}) gives a system generating flows connected with $t^2_n$,
\bea
&&
(z'-v_j)D^2(z')u_j-u_jD^2(z')v_j
\nn\\&&\qquad
=(S^1_x u_j)_y-2S^1_{xx}(z')
\sum_{i(i\neq j)}\frac{u_i}{v_j-v_i} +(u_jS^1_{xy}-u_{jx}S^1_y)
\nn\\&&
(z'-v_j)D^2(z')v_j=-S^1_{xx}u_j -(S^1_y v_{jx}-S^1_xv_{jy}).
\label{nred2}
\eea

Let us consider the first systems of the reduced hierarchy. The first order of expansion
of (\ref{linred1}), (\ref{linred2}) in $z'^{-1}$ provides linear equations for these systems
(plus (\ref{linred0})),
\beaa
&&
(\partial_t -z\partial_x)S^1=
\left(-\Theta_{xx}\sum_i\frac{u_i}{z-v_i} + \Theta_{xy}\right)\partial_x S^1(z),
\\&&
(\partial_{\tilde t} -z\partial_y)S^1=
\left(-\Theta_{xy}\sum_i\frac{u_i}{z-v_i} + \Theta_{yy}\right)\partial_x S^1(z).
\eeaa
The first order of expansion
of (\ref{nred1}), (\ref{nred2}) in $z'^{-1}$ gives the first systems of reduced hierarchy,
\bea
&&
(\partial_t-v_j\partial_x)u_j-u_j\partial_x v_j
\nn\\&&\qquad
=-(\Theta_{xx} u_j)_y+2\Theta_{xxx}
\sum_{i(i\neq j)}\frac{u_i}{v_j-v_i} -(u_j\Theta_{xxy}-u_{jx}\Theta_{xy})
\nn\\&&
(\partial_t-v_j\partial_x)v_j=\Theta_{xxx}u_j +(\Theta_{xy} v_{jx}-\Theta_{xx} v_{jy})
\label{nred10}
\eea
and
\bea
&&
(\partial_{\tilde t} -v_j \partial_y)u_j-u_j\partial_y v_j
\nn\\&&\qquad
=-(\Theta_{xy} u_j)_y+2\Theta_{xxy}
\sum_{i(i\neq j)}\frac{u_i}{v_j-v_i} -(u_j\Theta_{xyy}-u_{jx}\Theta_{yy})
\nn\\&&
(\partial_{\tilde t} -v_j \partial_y)v_j=\Theta_{xxy}u_j +(\Theta_{yy} v_{jx}-\Theta_{xy}v_{jy}),
\label{nred20}
\eea
where $\Theta$ is defined by relation (\ref{Thetared}).

Now we will consider some simple special cases of the general reduction,
when the function $S^1$ has simple analytic properties in $z$.
\subsection*{Logarithmic reduction}
In this case $S^1$ is of the form
$$
S^1=S^1_+ - \sum_{i=1}^Nc_i\ln(1-\frac{u_i}{z}).
$$
Generating equations for the reduced hierarchy read
\beaa
(z'-u_j)D^1(z')u_j=-\{D^1(z')\Theta, u_j\},
\\
(z'-u_j)D^2(z')u_j=-\{D^2(z')\Theta, u_j\},
\\
D^2(z')\Theta=\sum_{i=1}^Nc_i\ln(1-\frac{u_i}{z'}).
\eeaa
The first two (2+1)-dimensional systems of
reduced hierarchy are
\beaa
\partial_{\tilde t}u_k=u_k\partial_y u_k +\sum_i c_i \{u_i, u_k\}
\eeaa
and
\beaa
&&
\partial_{t}u_k=u_k\partial_x u_k - (u_k)_x \partial_x \sum_i c_i u_i -\Theta_{xx} \partial_y u_k,
\\
&&
\Theta_y=-\sum_i c_i u_i.
\eeaa
Common solution to these systems gives a solution $\Theta$ to heavenly
equation (\ref{heav1}).
\subsection*{Rational reduction}
We consider $S^1$ of the form
$$
S^1=S^1_+ + \sum_{i=1}^N\frac{u_i}{z-z_i}.
$$
Generating equations for the reduced hierarchy read
\beaa
(z'-z_j)D^1(z')u_j=-\{D^1(z')\Theta, u_j\},
\\
(z'-z_j)D^2(z')u_j=-\{D^2(z')\Theta, u_j\},
\\
D^2(z')\Theta=-\sum_{i=1}^N\frac{u_i}{z'-z_i}.
\eeaa
The first two (2+1)-dimensional systems of
reduced hierarchy are
\beaa
\partial_{\tilde t}u_k=z_k\partial_y u_k +\sum_i\{ u_i, u_k\}
\eeaa
and
\beaa
&&
\partial_{t}u_k=z_k\partial_x u_k - (u_k)_x \partial_x \sum_i u_i -\Theta_{xx} \partial_y u_k,
\\
&&
\partial_y \Theta=-\sum_i u_i.
\eeaa

\subsection*{(1+1)-dimensional reductions}
If we use rational or logarithmic reduction for both $S^1$, $S^2$,
we obtain (1+1) dimensional systems of equations for coefficients directly from (\ref{bracket}).
The reduction with both $S^1$, $S^2$ rational
was considered in \cite{Gindikin}.

Let us use logarithmic reduction for both $S^1$, $S^2$,
\beaa
S^1=S^1_+ - \sum_{i=1}^Nc_i\ln(1-\frac{u_i}{z}),\\
S^2=S^2_+ - \sum_{i=1}^Mc_i\ln(1-\frac{v_i}{z}).
\eeaa
Then from (\ref{bracket}) we get a (1+1)-dimensional system of equations
\beaa
\partial_x u_k +\sum_i c_i \frac{\{u_k,v_i\}}{u_k-v_i}=0,
\\
\partial_y v_j -\sum_i c_i \frac{\{v_j,u_i\}}{v_j-u_i}=0.
\eeaa
Using the expressions
$$
S^1_n=\sum_{i=1}^N c_i\frac{u_i}{i},\quad S^2_n=\sum_{i=1}^M c_i\frac{v_i}{i},
$$
we obtain the systems defining the dependence of $u_k$, $v_j$ on higher times,
the first two of them are
\beaa
\left\{
\begin{array}{lcl}
\partial_{t}u_k&=&u_k\partial_x u_k -\sum_i c_i \{v_i, u_k\},\\
\partial_{t}v_j&=&v_j\partial_x v_j -\sum_i c_i \{v_i, v_j\},
\end{array}
\right.
\\
\left\{
\begin{array}{lcl}
\partial_{\tilde t}u_k&=&u_k\partial_y u_k +\sum_i c_i \{u_i, u_k\},\\
\partial_{\tilde t}v_j&=&v_j\partial_y v_j +\sum_i c_i \{u_i, v_j\}.
\end{array}
\right.
\eeaa

\section{Symmetry constraints}
In this section we will consider symmetries of the heavenly equation hierarchy
defined through the wave functions of the hierarchy (solutions to
linear equations of the hierarchy) and symmetry constraints connected
with these symmetries.
Symmetries of this type were discussed in the work
\cite{heav} starting from explicit formula for the function $\Theta$. Similar
symmetry constraints are well known in KP hierarchy case (see e.g. \cite{KStr1})
as well as in
dispersionless case \cite{dconstr}. We will demonstrate that symmetry constraint is
equivalent to rational reduction (for one of the functions $S^1$, $S^2$)
of the heavenly equation hierarchy.

We inroduce a set of wave functions $\sigma_i(\mathbf{t}_1,\mathbf{t}_2)$ depending only
on the times of the hierarchy (no dependence on $z$),
\bea
&&
(z-z_i)D^1(z)\sigma_i=-\{S^2(z),\sigma_i\}
\label{wave1}
\\
&&
(z-z_i)D^2(z)\sigma_i=\{S^1(z),\sigma_i\},
\label{wave2}
\eea
where $z_i$, $1\leqslant i\leqslant N$, is some fixed set of points.

\begin{theorem}
$\delta\Theta=\sigma_i$ is an infinitesimal symmetry for $\Theta$
(i.e., it satisfies linearized equations of the hierarchy).
\end{theorem}
\textbf{Proof}
Taking vertex cross-derivatives of (\ref{wave1}), (\ref{wave2}), we get
$$
\{S^1(z),D^1(z)\sigma_i\}+\{S^2(z),D^2(z)\sigma_i\}=0.
$$
Then, using the representation of $S^1, S^2$ in terms of $\Theta$ (\ref{repr}), we obtain
\beaa
D^2(z)\partial_x \sigma_i-D^1(z)\partial_y \sigma_i-
\{D^1(z)\sigma_i,D^2(z)\Theta\}-\{D^1(z)\Theta,D^2(z)\sigma_i\}=0,
\eeaa
that is exactly the linearization of equation (\ref{heav1}). In a similar manner,
it possible to prove
that $\sigma_i$ satisfies the linearization of a general set of addition formulae.
\hfill$\square$\\

Then it is possible to introduce the symmetry constraint
\be
\Theta_x=\sum_{i=1}^N \sigma_i.
\label{constr1}
\ee

\begin{theorem}
The constraint (\ref{constr1}) is equivalent to
\be
S^2(z)=S^2_+(z) +\sum_{i=1}^N \frac{\sigma_i}{z-z_i}.
\label{rational}
\ee
\label{ratl}
\end{theorem}
\textbf{Proof}
First, it is straightforward to demonstrate (using (\ref{repr}))
that constraint (\ref{constr1}) is a necessary condition
for $S^2$ to be of the form (\ref{rational}). To prove that it is sufficient,
we will prove first the uniqueness of $S^2$ satisfying the set of linear
equations associated with the heavenly equation hierarchy.
\begin{lemma}
If the function $s^2(z,\mathbf{t}^1,\mathbf{t}^2)$ satisfies linear equations
\bea
&&
(z'-z)D^1(z'){s^2}(z)=-\{S^2(z'),s^2(z)\}
\label{Gen1L}
\\
&&
(z'-z)D^2(z')s^2(z)=\{S^1(z'),s^2(z)\}
\label{Gen2L}
\eea
(or, equivalently, the set of linear equations associated with (\ref{Hi1}), (\ref{Hi2}))
and $(S^2)_+=(s^2)_+$,
then $s^2=S^2$ (up to a function of $z$ only).
\label{unique}
\end{lemma}
\textbf{Proof} (Lemma \ref{unique})  Taking (\ref{Gen1L}), (\ref{Gen2L}) at $z=z'$, we get
$$
\{S^2(z),s^2(z)\}=0,\quad \{S^1(z),s^2(z)\}=1.
$$
Taking into account that $\{S^1,S^2\}=1$, we come to the conclusion that
$$
s^2(z)=S^2_+(z) + \phi,
$$
where $\phi_x=\phi_y=0$. Substituting $s^2$ to (\ref{Gen1L}), (\ref{Gen2L}) and taking into account
that $\phi$ annulates the Poisson bracket, we obtain that
$$
D^1(z')\phi(z)=D^2(z')\phi(z)=0,
$$
thus $\phi$ is independent of all times of the hierarchy,
so it doesn't influence the dynamics and reflects a freedom in the definition of $S^2$.
\hfill$\square$\\

To finish the proof of Proposition \ref{ratl},
it is enough to demonstrate that under the constraint (\ref{constr1})
function in the r.h.s. of (\ref{rational}) satisfies
equations (\ref{Gen1L}), (\ref{Gen2L}). Substituting this function to
(\ref{Gen2L}), we get
$$
1+(z'-z)\sum_i\frac{1}{z-z_i}\frac{1}{z'-z_i}\{S^1(z'),\sigma_i\}=
{S^1_x(z')}+ \sum_i\frac{1}{z-z_i}\{S^1(z'), \sigma_i\}.
$$
Both l.h.s. and r.h.s. are rational in $z$, the coefficients of the poles at $z_i$ are evidently
equal, and the condition at $z=\infty$ is (we use $z$ instead of $z'$)
$$
(S^1_x)_-+\sum_1 \frac{1}{z-z_i}\{S^1,\sigma_i\}=0.
$$
Using the equations for $\sigma_i$ (\ref{wave1}), (\ref{wave2}) to get
$$
(S^1_x(z))_-+D^2(z)\sum_i \sigma_i=0,
$$
and applying the constraint (\ref{constr1}) we discover that the condition
at $z=\infty$ is indeed satisfied,
so the function $S^2_+(z) +\sum_{i=1}^N \frac{\sigma_i}{z-z_i}$ satisfies (\ref{Gen2L}).
In a similar manner,
it is possible to prove that this function satisfies (\ref{Gen1L}). Then, using Lemma \ref{unique},
we come to the conclusion that
$$
S^2(z)=S^2_+(z) +\sum_{i=1}^N \frac{\sigma_i}{z-z_i}.
$$
\hfill$\square$\\
Finally, we will formulate a more general statement; the proof is completely analogous.
\begin{theorem}
The constraint
\be
\partial^1_n\Theta=\sum_{i=1}^N \sigma_i.
\label{constr1gen}
\ee
is equivalent to
\be
S^2(z)=S^2_+(z) +\sum_{j=1}^{n}\frac{\nu_j}{z^j}+\frac{1}{z^n}\sum_{i=1}^N \frac{\sigma_i}{z-z_i},
\label{rationalgen}
\ee
where the functions $\nu_j$ are defined by the relations
$$
\partial^1_{j-1}\Theta=\nu_j.
$$
\label{ratlgen}
\end{theorem}
\subsection*{Acknowledgments}
LVB is grateful to Dipartimento di Fisica dell' Universit\`a di Lecce
and Sezione INFN
for hospitality and support; the visit was
partially supported
by the grant COFIN 2004 `Sintesi'.
LVB was also supported in part by RFBR grant 04-01-00508,
President of Russia grant 1716-2003 (scientific schools)
and RAS Presidium program  `Mathematical methods of nonlinear dynamics'.

BGK was supported in part
by the grant COFIN 2004 `Sintesi'.

\end{document}